\documentclass[aps,showpacs,floats,twocolumn,floatfix,superscriptaddress]{revtex4-1}
\setcitestyle{super}
\pdfoutput=1
\usepackage{lipsum}
\usepackage{mathrsfs}
\usepackage{bm,amsbsy,amssymb,amsmath}
\usepackage{caption}
\usepackage{subcaption}
\usepackage{graphics,graphicx,dcolumn,fleqn,epic,eepic,float,tabularx}
\usepackage{multirow,rotate,rotating,color}
\usepackage[utf8]{inputenc}
\newcommand{\figref}[1]{Fig.~\ref{fig:#1}}
\newcommand{\eqnref}[1]{Eq.~(\ref{eq:#1})} 
  \definecolor{tuered}{RGB}{214,0,74}
  \definecolor{tueblue}{RGB}{0,102,204}
  
  \newcommand{\revisedtext}[1]{\textcolor{black}{#1}}

\usepackage{tikz,pgfplots}
\usepackage{relsize}
\tikzset{fontscale/.style = {font=\relsize{#1}}}
\usetikzlibrary{calc}
\graphicspath{{img/}}

\usepackage[colorinlistoftodos,prependcaption,textsize=tiny]{todonotes}

\begin{document}
\title{The effect of the liquid layer thickness on the \\dissolution of immersed surface droplets}
 \author{Qingguang Xie}
  \email{q.xie1@tue.nl}
  \affiliation{Department of Applied Physics, Eindhoven University of Technology, P.O. Box 513, 5600MB Eindhoven, The Netherlands}
  \author{Jens Harting}
  \email{j.harting@fz-juelich.de}
  \affiliation{Helmholtz Institute Erlangen-N\"urnberg for Renewable Energy (IEK-11), Forschungszentrum J\"ulich, F\"urther Str. 248, 90429 N\"urnberg, Germany}
  \affiliation{Department of Applied Physics, Eindhoven University of Technology, P.O. Box 513, 5600MB Eindhoven, The Netherlands}
\begin{abstract}
Droplets on a liquid-immersed solid surface are key elements in many
applications, such as high-throughput chemical analysis and droplet-templated
porous materials. Such surface droplets dissolve when the surrounding liquid is
undersaturated and the dissolution process is usually treated analogous to a
sessile droplet evaporating in air. Typically, theoretical models predict the
mass loss rate of dissolving droplets as a function of droplet geometrical
factors (radius, constant angle), and droplet material properties (diffusion
constant and densities), where the thickness of the surrounding liquid layer is
neglected.  
Here, we investigate, both numerically and theoretically, the effect
of the liquid layer thickness on the dissolution of surface droplets. 
We perform $3D$ lattice Boltzmann simulations and obtain
the density distribution and time evolution of droplet height during dissolution.
Moreover, we find that the dissolution slows down and the lifetime linearly
increases with increasing the liquid layer thickness. 
We propose a theoretical model based on a quasistatic diffusion equation 
which agrees quantitatively with simulation results for thick liquid layers.
Our results offer insight to the
fundamental understanding of dissolving surface droplets and can provide
valuable guidelines for the design of devices where the droplet lifetime is of
importance.

\end{abstract}

\maketitle

\section{Introduction} 
 \label{sec:intro}
Droplets on a substrate immersed in a liquid film have practical implications
for a wide range of applications from biomolecular analysis and chemical
reactions in microfluidic devices to high-resolution imaging
techniques~\cite{Detlef2015,Mendez-Vilas2009,Chiu2009,Shemesh2014}. Such
surface droplets can be produced by the solvent exchange
method~\cite{ZhangPNAS2015,BaoLei2016}, microprinting~\cite{Dietrich2015},
emulsion direct adsorption~\cite{Xuehua2008} and others~\cite{Day2012}.

\begin{figure*}[t!]
\centering
 \begin{tikzpicture}
 \fill[blue,opacity=0.4] (-9,0) rectangle (-1.4,3.0);
 \fill[red,opacity=0.1] (-9,3.0) rectangle (-1.4,3.5);
 \draw[black, dashed] (-9,3) to (-4,3);
 \node at (-3.3,3.25) [fontscale=2] {ambient air};
 
 \filldraw[fill=red!50!white, draw=green!50!black]
      (-8.05,0) -- (-7.35,0.0) arc (0:180:0.7) -- (-8.05,0);
      \filldraw[fill=red!50!white, draw=green!50!black]
      (-6.15,0) -- (-5.45,0.0) arc (0:180:0.7) -- (-6.15,0);
       \filldraw[fill=red!50!white, draw=green!50!black]
      (-4.25,0) -- (-3.55,0.0) arc (0:180:0.7) -- (-4.25,0);
          \filldraw[fill=red!50!white, draw=green!50!black]
      (-2.35,0) -- (-1.65,0.0) arc (0:180:0.7) -- (-2.35,0);

  \draw[ultra thick,black] (-9,0) to (-1.4,0);
  
  \draw[thick, black,arrows=->] (-7.25,0) to (-7.25,3);
 \draw[thick, black,arrows=->] (-7.25,3) to (-7.25,0);
 \node at (-7.55,1.55) [fontscale=2] {$L_z$};
 
 \draw[thick, black,arrows=->] (-9,0) to (-9,1);
 \node at (-8.85,0.5) [fontscale=2] {z};
 
 \draw[black, dashed] (-6.15,0.0) to (-6.15,0.9);
 \draw[black, dashed] (-4.25,0.0) to (-4.25,0.9);
 \draw[thick,black, arrows=->] (-6.15,0.9) to (-4.25,0.9);
  \draw[thick,black, arrows=->] (-4.25,0.9) to (-6.15,0.9);
  \node at (-5.15,1.2) [fontscale=2] {$L$};
  
  \draw[thick, black,arrows=->] (-2.35,0) to (-2.35,0.7);
 \node at (-2.6,0.4) [fontscale=2] {h};
   \draw[thick, black,arrows=->] (-2.35,0.2) to (-1.65,0.2);
   \draw[thick, black,arrows=->] (-1.65,0.2) to (-2.35,0.2);
    \draw[black] (-1.65,0) to (-1.65,0.3);
 \node at (-2.1,0.4) [fontscale=1] {R};

\end{tikzpicture}
\caption{Sketch of a regular array of surface droplets
containing liquid $1$ sitting on a substrate and covered by a layer of another
liquid (liquid $2$) of thickness $L_z$. The maximum height of the droplets is
denoted by $h$ and the radius of their footprint is $R$.  $L$ is the center to
center distance between neighboring droplets. }
\label{fig:geo}
\end{figure*}
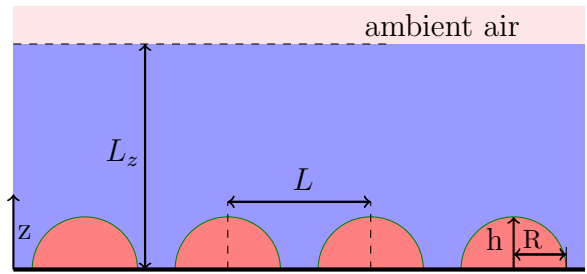

If the surrounding liquid is undersaturated with droplet liquid, the droplets
dissolve. 
The dissolution
process of surface droplets is similar to the dissolution of surface bubbles~\cite{Xiaojue2018,Lauga2018} 
and the evaporation of sessile droplets~\cite{Cazabat2010}.
There are two physical mechanisms that can affect the dissolution rate of a
surface droplet. The first mechanism is the rate at which liquid molecules
cross the droplet interface. The second mechanism is the transport of the
droplet liquid away from the droplet surface in the surrounding liquid.
Normally, the transfer rate of liquid molecules across the interface is much faster 
than the diffusion rate of the liquid~\cite{Cazabat2010}. 
Thus, the dissolution rate of the droplet is dominated by the diffusion of droplet liquid into the surrounding environment.
The diffusion of droplet liquid is driven by the gradient of the droplet liquid in the surrounding liquid, and 
the time-dependent density of droplet liquid in the surrounding liquid follows 
the unsteady diffusion equation, known as Fick's second law
\begin{align}
\partial \rho_1 / \partial t =  D \nabla^2 \rho_1,
\label{eq:unsteady}
\end{align}
where $D$ is the diffusion constant for droplet liquid in surrounding liquid and $\rho_1$ is the density of the droplet liquid.
Here, we denote the droplet liquid as liquid $1$ and the surrounding liquid as liquid $2$.
Due to dissolution,
the volume of the drop decreases until it is fully dissolved.  We
assume that the convective transport of liquid $1$ induced by the density
difference between liquid $1$ and liquid $2$ is negligible.
The diffusion timescale is characterized by $t_{df}\sim R^2/D$, where $R$ is the base radius of the droplets, as shown in~\figref{geo}. 
The dissolution timescale can be expressed as $t_{ds}\sim \rho^{1}_{d}/(\rho^{1}_{s}-\rho^{1}_{top})t_{df}$, 
where $\rho^{1}_{d}$ is the density of liquid $1$ inside the droplet, $\rho^{1}_{s}$ is the saturation density of liquid $1$ near the
droplet surface and $\rho^{1}_{top}$ is the density of liquid $1$ in the ambient air. 
The ratio between the dissolution timescale and diffusion timescale $t_{ds}/t_{df}\sim \rho^{1}_{d}/(\rho^{1}_{s}-\rho^{1}_{top})$ is typically much larger than $10^2$.
Therefore, the time-dependent term in~\eqnref{unsteady} can be neglected and the density of liquid $1$ follows
a quasi-steady diffusion equation $D \nabla^2 \rho_1 = 0$.
For the dissolution of droplet arrays, as shown in~\figref{geo}, the boundary conditions are: 
i) the density of liquid $1$ equals the saturation density along the droplet surface, 
$\rho^{1}_{r_k=R}=\rho^{1}_{s}$, where $r_k$ is the local radial coordinate of droplet $k$ ; 
ii) the liquid density at the liquid-air interface is the liquid density in the ambient air,
$\rho^{1}_{z=L_z}=\rho^{1}_{top}$; iii) the substrate is impermeable, $\partial \rho_{1} /\partial z=0$ along the substrate. 
There is no analytical solution available for the unsteady diffusion equation 
or even the quasistatic diffusion equation with the above boundary conditions. 

In the case where the liquid layer thickness $L_z$ is much larger than the droplet height, $L_z/h \rightarrow \infty$, 
and the inter-distance of droplets is much larger than the droplet base radius, $L/R \rightarrow \infty$, 
the system can be treated like a single dissolving surface droplet, which 
is analogous to a sessile droplet evaporating in an open air environment.
During evaporation of a droplet, the quasistatic diffusion
equation for the concentration field $\rho$, $\nabla^2 \rho = 0$ is then used
to predict the evaporation rate along the surface of the
droplet~\cite{Deegan1997,Deegan2000,Popov2005}. 
The boundary conditions become: i) the density of liquid $1$ equals the saturation 
density along the droplet surface, $\rho=\rho_{s}$; ii) the liquid density far away from the droplet is 
$\rho_{\infty}$; iii) the substrate is impermeable, $\partial \rho /\partial z=0$ along the substrate. 
Assuming the effect of gravity can be neglected, the droplet takes a spherical cap shape dominated by surface tension.
In this case, the evaporation flux $J=-D\nabla \rho$ is analogous to the electric potential distribution around a charged 
lens-shaped conductor and an analytical solution was derived by Lebedev~\cite{Lebedev}. 
Popov~\cite{Popov2005} used this analytical solution and obtained the total mass flux 
by integrating the evaporation flux over the droplet
surface,
\begin{align}
 &\frac{dM}{dt} =-\pi R D (\rho_s - \rho_\infty) \nonumber \\
 &\left(\frac{\sin \theta}{1+\cos \theta} + 4\int_0^\infty \frac{1+ \cosh 2\theta \epsilon}{ \sinh 2 \theta \epsilon} \tanh \left[(\pi - \theta)\epsilon \right] d\epsilon\right) 
 \mbox{,} 
 \label{eq:evaporate}
\end{align}
where $\theta$ is the contact angle of the droplet.

Dietrich et al.~\cite{Dietrich2015,Laghezza2016} applied the quasistatic diffusion approach and~\eqnref{evaporate}
to study the dissolution of single and multiple $1$-heptanol or $3$-heptanol surface droplets 
in a container filled with undersaturated water. They find good
agreement between experimental data and the above equation in the limit where
the surrounding water is still highly undersaturated after all the droplets are
dissolved~\cite{Dietrich2015,Dietrich2016,Laghezza2016}. However, different from
evaporating droplets in an open air environment, the surrounding liquid film
during dissolution of surface droplets can be saturated while the surface
droplets are still present.  Then, the dissolution rate of the droplets is
determined by the diffusion rate of their liquid in the liquid environment and
the distance the droplet liquid has to travel to reach the ambient air
(see~\figref{geo}).  In this case, the total mass flux is determined by
integrating the diffusive flux along the free surface where the liquid meets
the ambient air. Thus, the layer thickness of the surrounding liquid is
expected to play an important role on the dissolution rate and the lifetime of dissolving surface
droplets.

In the remainder of this paper, we numerically investigate the effect of the
layer thickness on the dissolution and the lifetime of dissolving surface droplets. We carry out
$3D$ lattice Boltzmann simulations with a diffusion dominated dissolution
model. A substrate is located at the bottom of the system and a surface droplet
is deposited on the substrate covered by another liquid layer of thickness
$L_z$. 
Periodic boundary conditions are applied at the sides of the system, to mimic an
infinite system with uniformly distributed droplets.  We initially saturate the
liquid layer and then apply the dissolution boundary condition at the top of
the system.  Different modes of contact line dynamics during dissolution such
as constant angle (CA) and constant radius (CR) modes are studied. We find
that the droplet lifetime is proportional to the layer thicknesses $L_z$.  The
simulations are accompanied by a theoretical analysis based on a quasistatic
diffusion equation which confirms the simulation result.

\section{Simulation Method}
\label{sec:method}
Our simulations are based on the lattice Boltzmann method (LBM) which can be
seen as an alternative way to approximate solutions of the Navier-Stokes
equations~\cite{Succi2001} and which  was demonstrated to be a powerful tool to
simulate multiphase/multicomponent fluids~\cite{Succi2001,Shan1993,Liu2016}. 
We use the pdeudopotential multicomponent LBM proposed by Shan and Chen~\cite{Shan1993},
which has been successfully applied to a wide range of multiphase/multicomponent 
flow problems during the past two decades~\cite{CHEN2014210}.
A number of groups have simulated problems
related to the diffusion or evaporation of fluids using the LBM recently.
Ledesma-Aguilar et
al.~\cite{LedVelYeo14,AguVelYeo16} present a diffusion based evaporation method based on the free energy multiphase lattice Boltzmann method and demonstrate quantitative agreement with
several benchmark cases as well as qualitative agreement with experiments
involving evaporating droplet arrays.
Jansen et al.~\cite{Jansen2013} study the evaporation of droplets on a chemically patterned
substrate and qualitatively compare the simulation results with experimental
data. Our group recently
applied the LBM together with the multicomponent method of Shan and Chen successfully to study the evaporation of a planar film, a
floating droplet, a sessile droplet, and a colloidal suspension droplet~\cite{DennisXieJens2016,XH18a}.
In the following we review some details of the method and refer the reader to
the relevant literature for a more detailed description and our
implementation~\cite{HyvKunHarH11,Jansen2011,Frijters2012,Gunther2013,Xie2016,
DennisXieJens2016,Liu2016,XH18a}.

In our implementation, two fluid components $c=1,2$ follow the evolution of their individual distribution functions 
discretized in space and time,
\begin{eqnarray}
  \label{eq:LBG}
  &&f_i^c(\mathbf{x} + \mathbf{e}_i \Delta t , t + \Delta t)-f_i^c(\mathbf{x},t) = \nonumber \\
  &&- \frac{\Delta t} {\tau^c} \left[  f_i^c(\mathbf{x},t) -f_i^\mathrm{eq}(\rho^c(\mathbf{x},t),\mathbf{u}^c(\mathbf{x},t))\right] 
    \mbox{,}
\end{eqnarray}
where $i=1,...,19$. $f_i^c(\mathbf{x},t)$ are the single-particle distribution
functions for each fluid component and $\mathbf{e}_i$ is the discrete
velocity in the $i$th direction. $\tau^c$ is the relaxation time for component
$c$. We define the macroscopic densities and velocities for
each component as $ \rho^c(\mathbf{x},t) = \rho_0
\sum_if^c_i(\mathbf{x},t)$, where $\rho_0$ is a reference density, and
$\mathbf{u}^c(\mathbf{x},t) = \sum_i  f^c_i(\mathbf{x},t)
\mathbf{e}_i/\rho^c(\mathbf{x},t)$, respectively. 
Here, $f_i^\mathrm{eq}$ is a second-order equilibrium distribution function~\cite{Qian1992}, defined as
\begin{align}
  \label{eq:eqdis}
  f_i^{\mathrm{eq}}(\rho^c,\mathbf{u}^c) = \omega_i \rho^c \bigg[ 1 + \frac{\mathbf{e}_i \cdot \mathbf{u}^c}{c_s^2} - 
  \frac{ \left( \mathbf{u}^c \cdot \mathbf{u}^c \right) }{2 c_s^2} + \frac{ \left( \mathbf{e}_i \cdot \mathbf{u}^c \right)^2}{2 c_s^4}  \bigg] \mbox{.}
\end{align}
$\omega_i$ is a coefficient depending on the direction: $\omega_0=1/3$
for the zero velocity, $\omega_{1,\dots,6}=1/18$ for the six nearest neighbors
and $\omega_{7,\dots,18}=1/36$ for the nearest neighbors in diagonal direction.
$c_s = \frac{1}{\sqrt{3}} \frac{\Delta x}{\Delta t}$ is the speed of sound.
When sufficient lattice symmetry is guaranteed, the Navier-Stokes equations can be recovered 
from \eqnref{LBG} on appropriate length and time scales~\cite{Succi2001}.
For convenience we choose the lattice constant $\Delta x$, the timestep $
\Delta t$, the unit mass $\rho_0 $ and the relaxation time $\tau^c$ to be
unity in the remainder of this article, 
which leads 
to a kinematic viscosity $\nu^c$ $=$ $\frac{1}{6}$ in lattice units. 
We note that the conversion from lattice units to physical units can be 
performed e.g. by matching dimensionless numbers, such as 
the Reynolds number or the Schmidt number~\cite{Timm2016}.

Following the work of Shan and Chen~\cite{Shan1993}, we apply a mean-field interaction force
\begin{align}
  \label{eq:sc}
 \mathbf{F}^c(\mathbf{x},t) = -\Psi^c(\mathbf{x},t) \sum_{\bar{c}} \sum_{i} \omega_i g_{c\bar{c}} \Psi^{\bar{c}}(\mathbf{x}+\mathbf{e}_i,t) \mathbf{e}_i
\end{align}
between fluid components $c$ and $\bar{c}$, in which $g_{c\bar{c}}$ is a constant interaction parameter. 
Here, $\Psi^c(\mathbf{x},t)$ is chosen as the functional form
 $ \Psi^c(\mathbf{x},t) \equiv \Psi(\rho^c(\mathbf{x},t) ) = 1 - e^{-\rho^c(\mathbf{x},t)}$.
We apply this force $\mathbf{F}^c(\mathbf{x},t)$ to the component $c$ by adding a 
shift $\Delta \mathbf{u}^c(\mathbf{x},t) =\frac{\tau^c \mathbf{F}^c(\mathbf{x},t)}{\rho^c(\mathbf{x},t)}$ 
to the velocity $\mathbf{u}^c(\mathbf{x},t)$ in the equilibrium distribution.

Inspired by the work of Huang et al., an interaction force is introduced between the fluid and the substrate~\cite{Huang2007}, 
\begin{align}
\mathbf{F}^c(\mathbf{x}) =  - g_{wc} \Psi^c (\mathbf{x}) \sum_{i} \omega_{i} s(\mathbf{x} + \mathbf{e}_{i}) \mathbf{e}_{i}
\mbox{,}
\label{eq:rhowall2}
\end{align}
where $g_{wc}$ is a constant. Here, $s(\mathbf{x} + \mathbf{e}_{i})= 1$ if
$\mathbf{x} + \mathbf{e}_{i}$ is a solid lattice site, and $s(\mathbf{x} +
\mathbf{e}_{i})=0$ otherwise.
An approximate formula
can be used to estimate the contact angle $\theta$ of a droplet on the substrate~\cite{Huang2007}:
\begin{align}
 \cos (\theta) = \frac{g^{wc}-g^{w\bar{c}}}{g^{c\bar{c}}[\Psi (\rho^c)-\Psi(\rho^{\bar{c}})]/2}
 \label{eq:huang}
\end{align}

The phase separation can be triggered by 
choosing a proper interaction parameter $g_{c\bar{c}}$ in~\eqnref{sc}.
Each component separates into a denser majority phase of density
$\rho_{ma}$ and a lighter minority phase of density $\rho_{mi}$,
respectively~\cite{Shan1993}. To simulate dissolution, we impose the distribution function of
component $c$ at the boundary sites $\mathbf{x}_H$ as~\cite{DennisXieJens2016}
\begin{align}
    f_i^c(\mathbf{x}_H,t) = f_i^\mathrm{eq}\left(\rho_H^c,\mathbf{u}^c_H(\mathbf{x}_H,t)\right),
\end{align}
in which $\mathbf{u}^c_H(\mathbf{x}_H,t)=0$.
Furthermore, for simplicity, we ensure total mass conservation within the system by 
setting the density of component $\bar{c}$ as
\begin{align}
    \rho^{\bar{c}}(\mathbf{x}_H,t) = \rho^c(\mathbf{x}_H,t-1) + \rho^{\bar{c}}(\mathbf{x}_H,t-1) - \rho^c_H,
\end{align}
so that the distribution functions of component $\bar{c}$ at the dissolution boundary sites $\mathbf{x}_H$ become
\begin{align} 
    f_i^{\bar{c}}(\mathbf{x}_H,t) = f_i^\mathrm{eq}\left(\rho_H^{\bar{c}},\mathbf{u}^{\bar{c}}_H (\mathbf{x}_H,t)\right),
\end{align}
where $\mathbf{u}^{\bar{c}}_H(\mathbf{x}_H,t)=0$.
When the imposed density
$\rho_H^c$ is lower than the equilibrium minority density $\rho_{mi}^c$, a
density gradient develops in the lighter minority phase of component $c$. This gradient
drives component $c$ to diffuse towards the dissolution boundary, which recovers the unsteady diffusion equation~\eqnref{unsteady} validated in our previous work~\cite{DennisXieJens2016}.
In the case that the densities of two components are similar, the bouyancy-driven convective flow can be neglected, and the dissolution is diffusion dominated and the diffusivity is given by~\cite{DennisXieJens2016}
\begin{align}
 D^c=\biggl[c_s^2(\tau-\frac{1}{2}) 
         -  \frac{c_s^2}{\rho^c+\rho^{\bar{c}}} (\rho^{\bar{c}} \Psi^c g_{c\bar{c}} \Psi^{'\bar{c}} + \rho^{c} \Psi^{\bar{c}}  g_{\bar{c}c} \Psi^{'c} )\biggr],
\end{align}
where $\Psi^{'c}$ and $\Psi^{'\bar{c}}$  are the spatial derivative of $\Psi^{c}$ and $\Psi^{\bar{c}}$, respectively.
A similar approach was recently introduced by Ledesma-Aguilar et al. for a free energy lattice Boltzmann method~\cite{LedVelYeo14}.
\begin{figure}[h]
\centering
 \includegraphics[width = 0.5\textwidth]{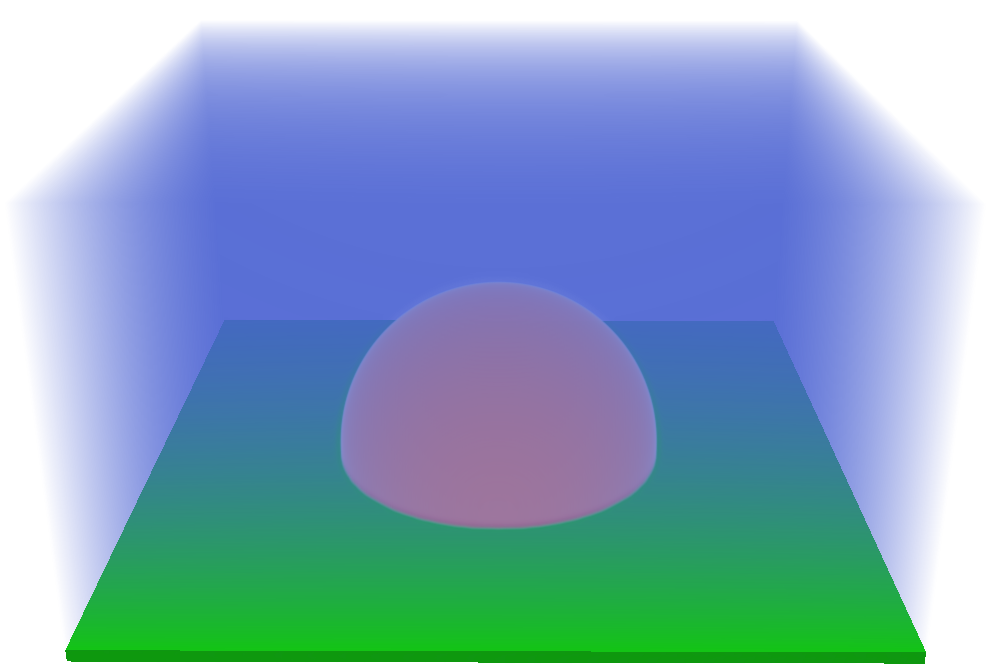}
 \caption{Snapshot of a surface droplet (red) sitting on a liquid (blue) 
 immersed substrate (green) obtained from our simulations.
 The system size is $256\times 256 \times 144$.  We apply 
 the dissolution boundary at the top plane and periodic boundary conditions at the sides of the system.
 }
 \label{fig:geo-simu}
\end{figure}

\section{Results and discussion}
\begin{figure}[h]
\begin{subfigure}{.23\textwidth}
\includegraphics[width= 0.99\textwidth]{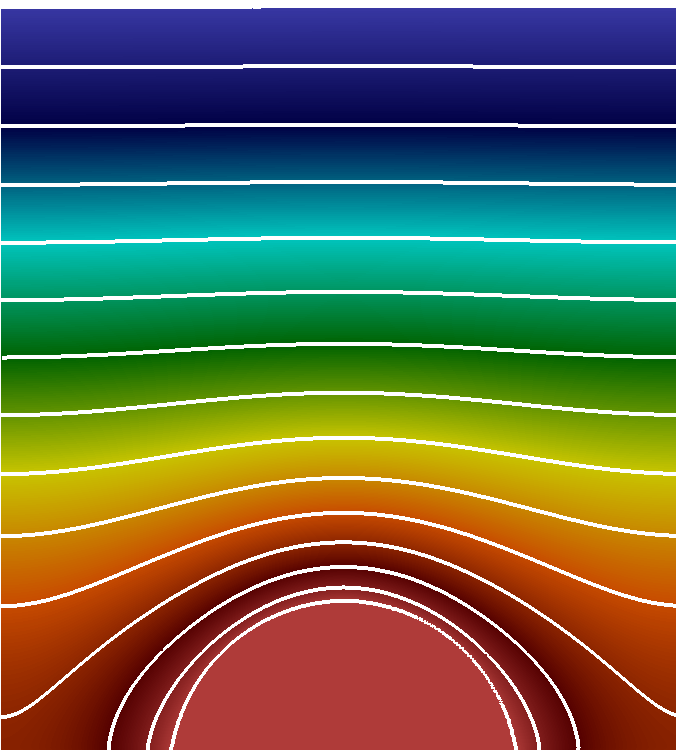}
 \subcaption{$t/T=0.0$}
 \label{fig:frs-1}
 \end{subfigure}
 \begin{subfigure}{.23\textwidth}
\includegraphics[width= 0.99\textwidth]{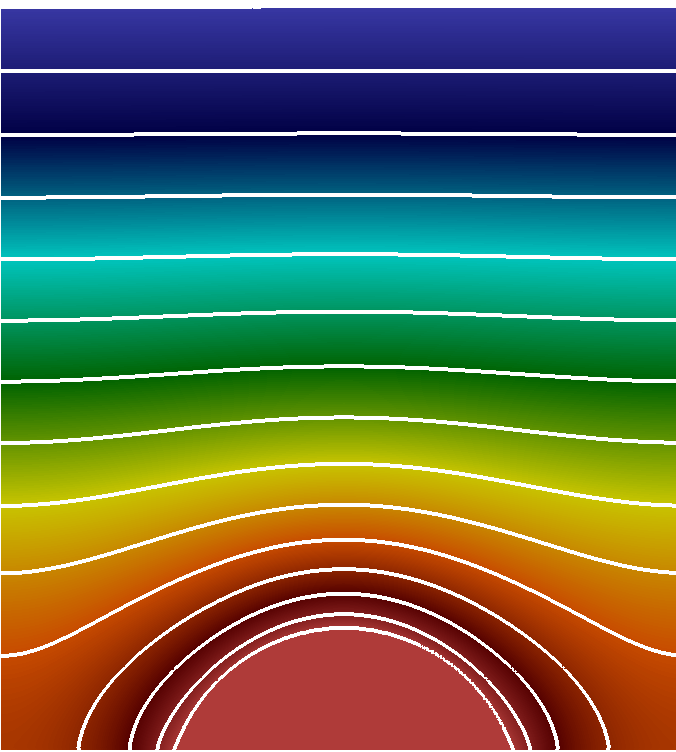}
 \subcaption{$t/T=0.28$}
 \label{fig:frs-2}
 \end{subfigure}
 \\
 \begin{subfigure}{.23\textwidth}
\includegraphics[width= 0.99\textwidth]{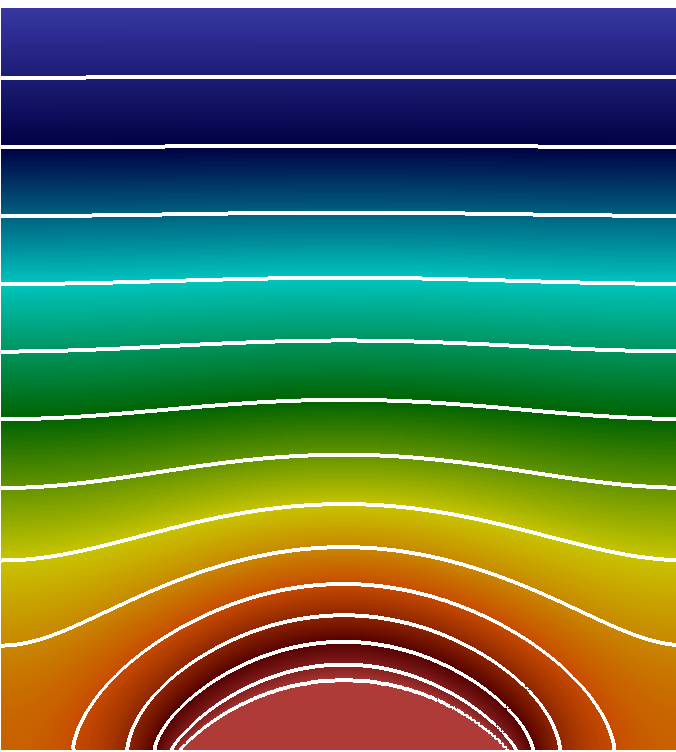}
 \subcaption{$t/T=0.57$}
 \label{fig:frs-3}
 \end{subfigure}
  \begin{subfigure}{.23\textwidth}
\includegraphics[width= 0.99\textwidth]{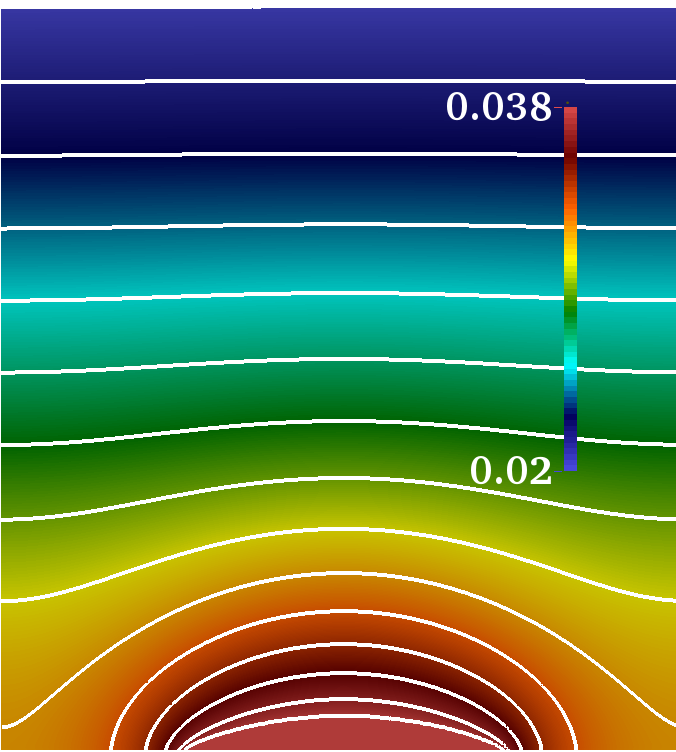}
 \subcaption{$t/T=0.85$}
 \label{fig:frs-4}
 \end{subfigure}
 \caption{Time evolution of the density distribution of liquid $1$ during 
 droplet dissolution obtained from our simulation.
The color represents the density and the white solid lines are iso-density lines. The droplet dissolves in constant radius mode. 
The position of the top surface corresponds to the top of the pictures.}
\label{fig:rhocr}
 \end{figure}

We simulate an infinite array of immersed droplets on a substrate as depicted
in~\figref{geo}. A snapshot from a simulation of a 3D unit cell is shown
in~\figref{geo-simu}.  The system size is $L \times L \times L_z$, where
$L=256$ and $L_z$ is chosen to be $L_z=72$, $144$, $288$, $576$, respectively.
We initialize the system with a droplet containing liquid $1$ of density
$\rho^{1}_{ma}=0.7$ and liquid $2$ of density $\rho^{2}_{mi}=0.04$. The
surrounding volume consists of liquid $2$ of density $\rho^{2}_{ma}=0.7$ and
liquid $1$ of density $\rho^{1}_{mi}=0.04$.  The droplet has an initial radius
$R_0=60$ and an initial contact angle $\theta_0=90^{\circ}$ corresponding to an
initial maximum height $h_0=60$.  The interaction strength in~\eqnref{sc} is
chosen to be $g_{12} = 3.6$ leading to a diffusivity $D\approx 0.12$. 
\revisedtext{The diffusion constant of the surrounding liquid in the droplet liquid is of similar order as the diffusion constant of the droplet liquid in the surrounding liquid.}
Furthermore, periodic boundary
conditions are applied at the sides of the system to mimic an infinite array of
identical droplets. After equilibration, we impose the dissolution boundary
condition at the top of the system to mimic the ambient air. Our dissolution
boundary condition allow to set a constant density and a zero velocity at the
top boundary~\cite{DennisXieJens2016}. The droplet liquid diffuses from the
droplet to the dissolution boundary, gradually forming a density gradient. 
We note that we only apply dissolution boundary conditions at the top
surface. Therefore, due to mass conservation the total diffusive flux
integrated along the droplet surface is consistent with the corresponding total
flux at the top surface.
The measurement of the lifetime of the droplet is only started once the gradient is
fully developed. We investigate the effect of layer thickness $L_z$ by varying
only the system size in $z$ direction and keeping all other parameters
constant.
 \begin{figure}[]
\centering
 \includegraphics[width = 0.5\textwidth]{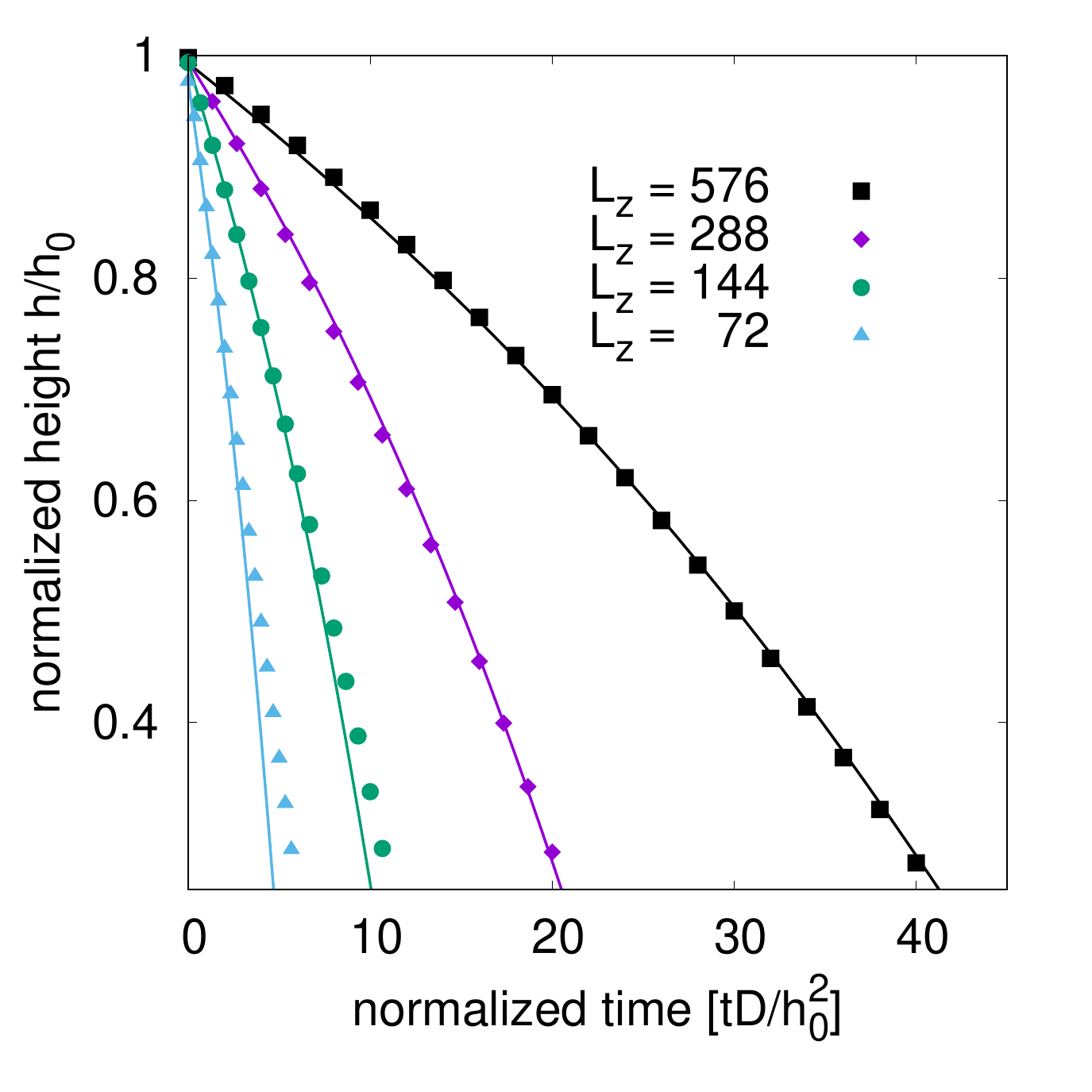}
 \caption{Time evolution of the droplet height during dissolution in CR mode. 
 The symbols are simulation data for different system heights length $Lz=74$ (triangles), $144$ (circles), $288$ (diamonds), and $576$ (squares), respectively.
 The solid lines denote the theoretical prediction~\eqnref{dhdt}.
}
 \label{fig:htcr}
\end{figure}

We start out our investigation with a dissolving droplet featuring a pinned
contact line, the so-called constant radius (CR) mode. The substrate is
chemically patterned with variable wettability: a superhydrophilic circle
($\theta \approx 0^{\circ}$) of radius $R_s = 60$ is located at the center
surrounded by a superhydrophobic area ($\theta \approx 180^{\circ}$).

\figref{rhocr} shows the time evolution of the density distribution of liquid
$1$ during the dissolution for a system of size $256\times 256 \times 288$.
Near the top of the system, the density is homogeneous along the horizontal
direction, while it follows the shape of the droplet in the vicinity of its
surface. The density gradient is larger near the droplet surface than at
a position far away from it. The dissolution flux is represented by the
distance between the neighboring iso-density lines. In the early state
(\figref{frs-1} and~\figref{frs-2}), the dissolution flux near the contact
line is smaller than that near the top of the droplet. This is due to the
collective effect introduced by neighboring droplets~\cite{Laghezza2016} and in
our case is an effect of the periodic boundary conditions in the horizontal
directions. \revisedtext{We note that this collective effect would become weaker when the droplet inter-distance increases, as observed in experiments~\cite{Lei2018,Laghezza2016}} 
 In the later stage (\figref{frs-3} and~\figref{frs-4}) when the
contact angle is much smaller than $90^{\circ}$, the dissolution flux diverges
towards the contact line, which is consistent with the theoretical prediction
of Popov~\cite{Popov2005}. In addition, the density gradient at the top of the
system stays almost constant during the dissolution. This indicates that the
effect of the curved droplet surface on the density distribution in the far
field is negligible if the thickness of liquid layer $L_z$ is much larger than the droplet height $h_0$, $L_z \gg
h_0$.

In~\figref{htcr} we show the simulation data (symbols) of the droplet height
versus time for different system heights $Lz=74$ (triangles), $144$ (circles), $288$
(diamonds), and $576$ (squares), respectively. 
We note that the multicomponent model of Shan and Chen suffers from spurious
vaporisation effects once the diameter of the droplets becomes $\approx5-10$
lattice units. To avoid the effect of the spurious vaporisation on the
analysis and to ensure a sufficient resolution, we only use simulation data
for droplet heights larger than $20$. \revisedtext{We note that the upper limit of droplet size is only given by the available computational resources. The method is furthermore valid as long as the droplets follow the continuum Navier Stokes assumptions and for sizes large enough so that thermal fluctuations do not play a role anymore.}
During the dissolution, the droplet height keeps decreasing, but with a slower rate when
the length $L_z$ is increased. This is reasonable because liquid $1$ requires more 
time to diffuse through a thicker liquid layer to arrive at the top free
surface.  Moreover, the decreasing rate of the droplet height speeds up towards
the end of the lifetime of the droplet.
\begin{figure}[]
\begin{subfigure}{.23\textwidth}
\includegraphics[width= 0.99\textwidth]{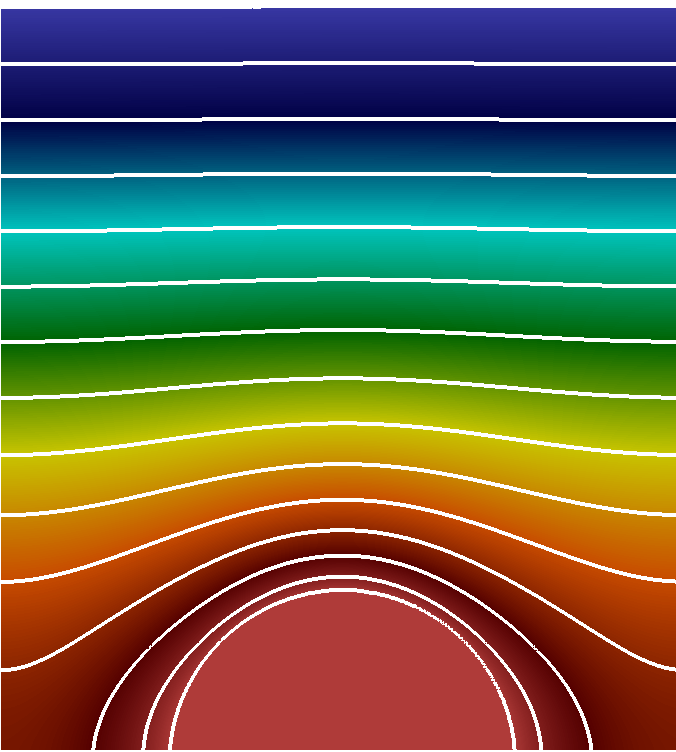}
 \subcaption{$t/T=0.0$}
 \label{fig:fca-1}
 \end{subfigure}
 \begin{subfigure}{.23\textwidth}
\includegraphics[width= 0.99\textwidth]{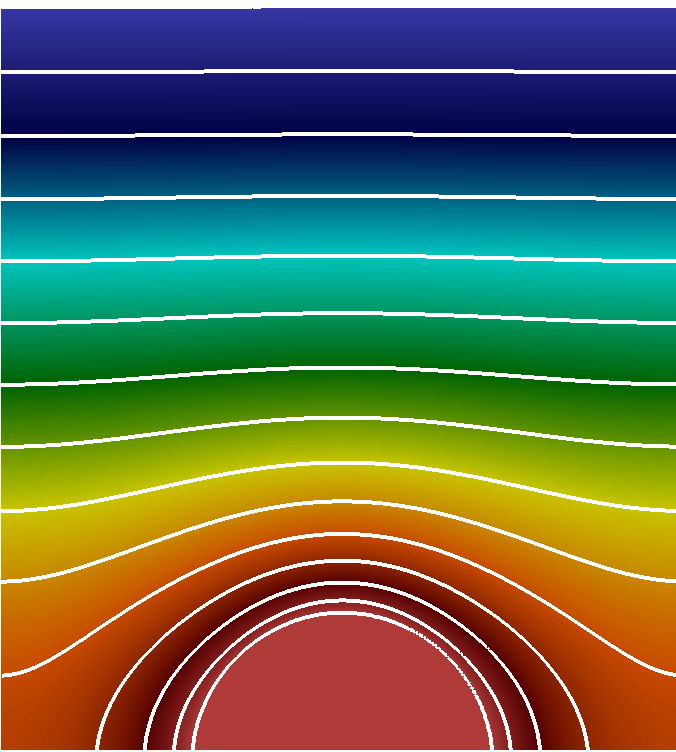}
 \subcaption{$t/T=0.31$}
 \label{fig:fca-2}
 \end{subfigure}
 \\
 \begin{subfigure}{.23\textwidth}
\includegraphics[width= 0.99\textwidth]{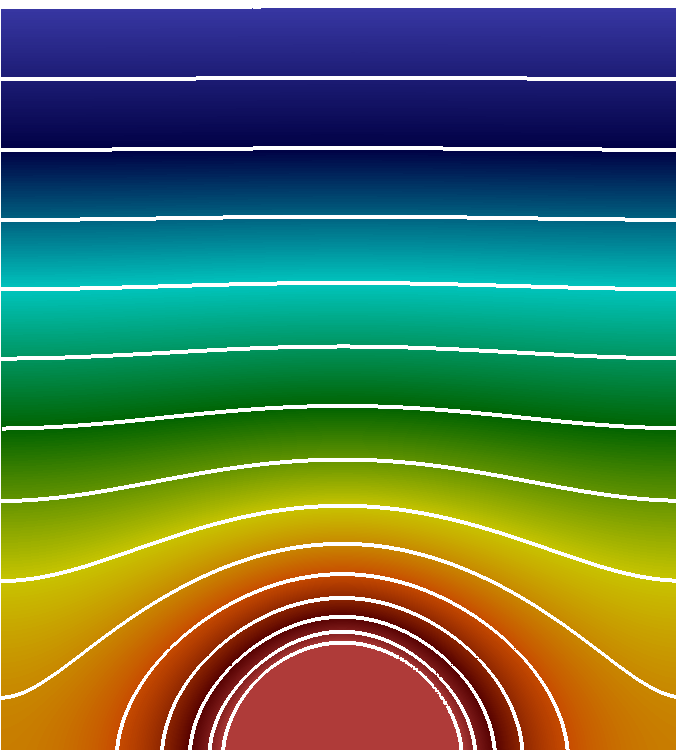}
 \subcaption{$t/T=0.62$}
 \label{fig:fca-3}
 \end{subfigure}
  \begin{subfigure}{.23\textwidth}
\includegraphics[width= 0.99\textwidth]{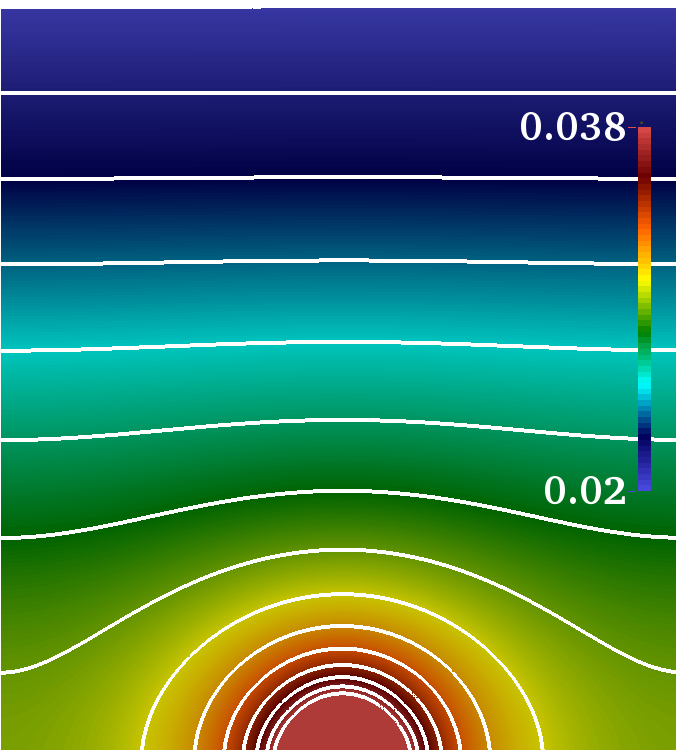}
 \subcaption{$t/T=0.93$}
 \label{fig:fca-4}
 \end{subfigure}
 \caption{Time evolution of the density distribution of liquid $1$ 
 during droplet dissolution obtained in our simulations. 
The color represents the density and the white solid lines are iso-density lines. The droplet dissolves in CA mode. 
The position of the top surface corresponds to the top of the pictures.}
\label{fig:rhoca}
 \end{figure}
 
\begin{figure}[h]
 \centering
 \includegraphics[width = 0.5\textwidth]{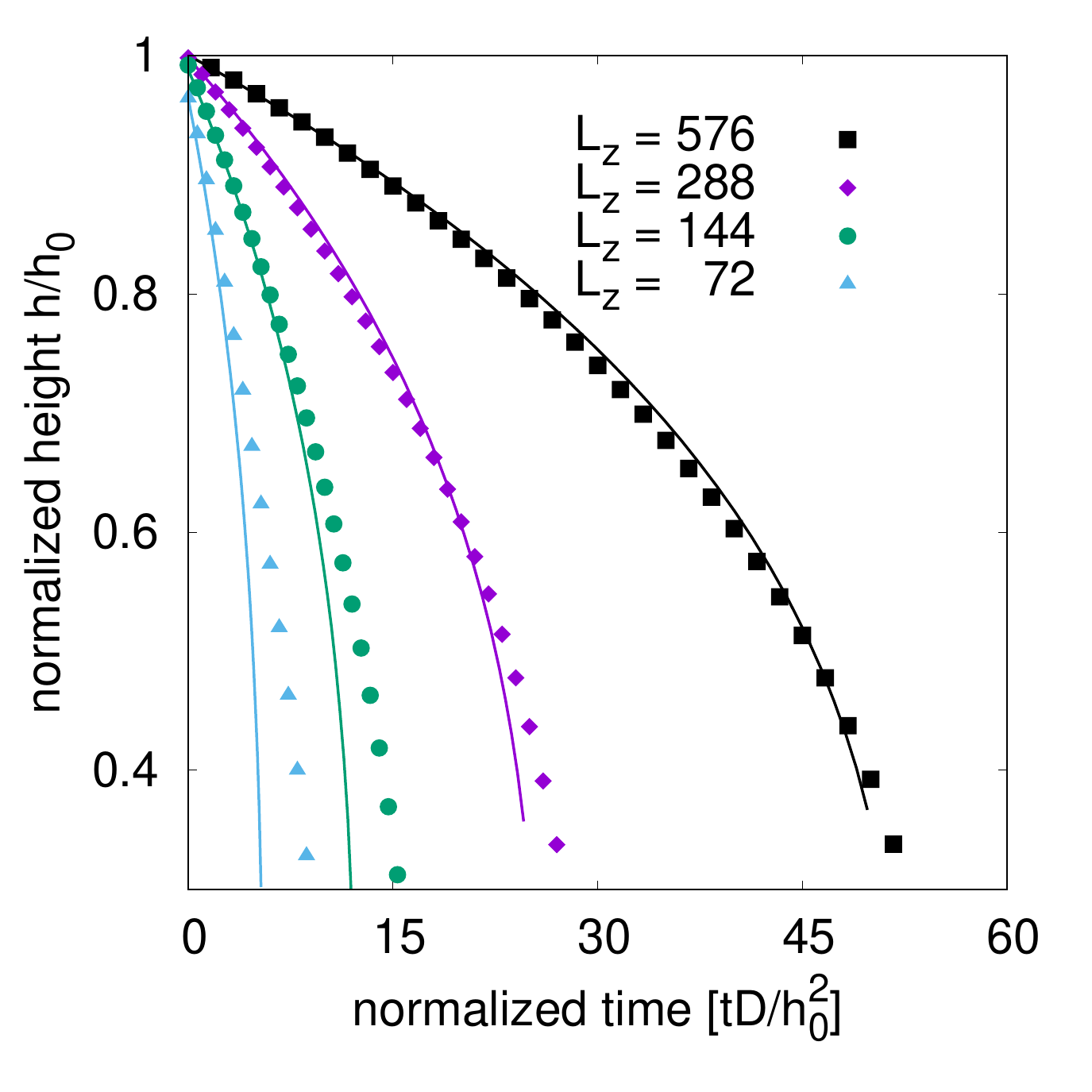}
 \caption{Time evolution of the droplet height during dissolution in CA mode. 
 The symbols are simulation data for different system heights $Lz=74$ (triangles), $144$ (circles), $288$ (diamonds) and $576$ (squares), respectively.
 The solid lines follow the theoretical prediction given by~\eqnref{ca}.}
 \label{fig:htca}
\end{figure}

Next, we investigate the dissolution of a droplet in constant angle (CA) mode. 
The substrate has a uniform wettability and the droplet keeps its contact angle of $90^{\circ}$ during the dissolution. \revisedtext{We note that in some cases, if the surrounding liquid diffuses into the droplet, the surrounding droplet molecules may change the chemical potential,
which affects the motion of the droplet and as such the contact angle
variation~\cite{Zhao2019}. However, in our system, the droplet is assumed to be saturated with the surrounding liquid and thus the diffusion of surrounding
liquid into the droplet is prohibited.}
In \figref{rhoca} we show the variation of the density distribution of liquid $1$ during 
dissolution for a system size of $256\times 256 \times 288$. 
Equivalent to the CR mode, the density distributes uniformly along the horizontal direction 
in the far field, whereas it is strongly affected 
by the curved droplet surface in the near field of the droplet.
The density gradient increases when approaching the droplet surface. 
Again, the collective effect introduced by neighboring droplets on the density gradient is observed in the 
earlier stages (\figref{fca-1} and~\figref{fca-2}).
This collective effect become weaker when the droplet radius deceases
(\figref{fca-3}) (i.e., the inter-spacing between neighboring droplets
increases), which is consistent with experimental
results~\cite{Laghezza2016,Lei2018}.
In the very late stage (\figref{fca-4}), the density gradient and thus the dissolution flux is uniform along the droplet surface, 
which is in agreement with the theoretical prediction of the evaporation flux along the surface of a droplet with $90^{\circ}$ contact 
angle~\cite{Popov2005}.

In~\figref{htca} we show the time evolution of droplet height obtained in our simulations (symbols) for 
different lengths $Lz=74$ (triangles), $144$ (circles), $288$ (diamonds) and $576$ (squares), respectively.
Similar to the CR mode, the droplet height decreases more slowly with increasing length $L_z$.
Additionally, the droplet shrinks faster towards the end of its lifetime.
\revisedtext{
We note that the convective flow induced by the movement of the droplet
interface during dissolution may have an effect on the effective dissolution
rate~\cite{YANG2018}.  To quantify the ratio of the contributions to mass transport by convection to
those by diffusion, we calculate the P{\'e}clet number $Pe=lu/D$, where $l$ is
a characteristic length, $u$ is a characteristic velocity and $D$ is the
diffusion constant.  Here, we have $u \sim 10^{-5}$, $D\sim 0.12$ and $l\sim 60
$ in lattice Boltzmann units, and obtain $Pe \sim 10^{-3}$. The P{\'e}clet
number is much smaller than $1$, therefore, the effect of convection near the
droplet surface on the mass transfer is negligible.  As discussed in the work
of Zhao et al.~\cite{YANG2018}, the interfacial shape can be affected by
the surface tension, viscosity and inertia during dissolution. In our system,
the Reynolds number $Re = l u / \nu^c$ is of the order of $10^{-3}$ and thus we
expect the effect of fluid inertia also to be negligible, i.e. the droplet
adheres to a spherical cap shape.
}

\begin{figure}[]
\centering
 \includegraphics[width = 0.5\textwidth]{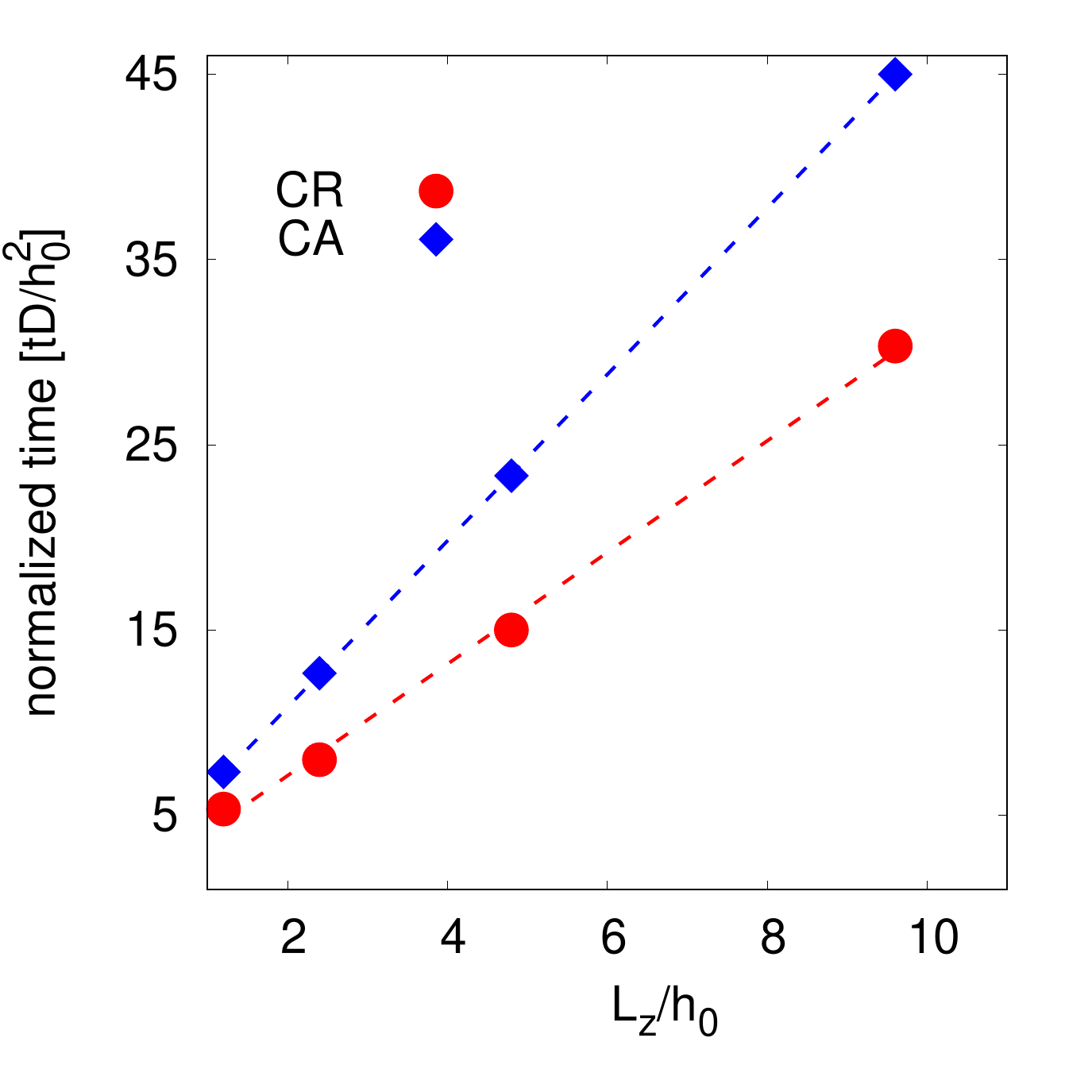}
 \caption{The normalized lifetime $tD/h_{0}^{2}$ of the dissolving surface droplet 
 for systems with different height $L_z/h_0$.  
 The symbols are simulation data for the CR mode (circles) and CA the mode (diamonds). The dashed lines are fitted linear functions, 
 indicating that the lifetime is to a good approximation proportional to the layer thickness $L_z$.}
 \label{fig:scale}
\end{figure}

In~\figref{scale}, we compare the lifetime $T$ of the dissolving surface droplet with increasing the layer thickness $L_z$
for both CR (circles) and CA (diamonds) modes. The dashed lines represent fitted linear functions.
We reiterate that the time zero in our measured lifetime corresponds to a moment after the density 
gradient is fully developed in the system and the droplet maximal height just begins to decrease. 
The results demonstrate that the lifetime of the dissolving surface droplet 
is strongly dependent on the thickness of the liquid layer $L_z$, i.e.
it increases approximately linearly with increasing height $L_z$ for both CR and CA mode. 

In the following we propose a theoretical model for taking into account the
effect of the layer thickness of the surrounding liquid on the lifetime of a
dissolving surface droplet.  We consider surface droplets of liquid $1$ sitting
on a substrate covered by liquid $2$ of thickness $L_z$, as illustrated
in~\figref{geo}. 
The droplets are uniformly distributed on the substrate with center to center
distance $L$ and are of identical size with initial maximum height $h_0$, and
initial radius of the contact line $R_0$.  We assume that the droplet is
surface tension dominated and thus adheres the shape of a spherical cap. 

Assuming a diffusion dominated dissolution and the buoyancy-driven convection
being negligible, the density of liquid $1$ follows a
quasi-static diffusion equation
\begin{align}
\partial_{t} \rho_1 = D \nabla^2 \rho_1 = 0. 
\end{align}
In the case of $L_z\gg h$, we can assume that the density of the droplet liquid
along the horizontal direction far away from the droplet surface is
homogeneous.  Furthermore, we can assume that the density varies linearly along the
$z$ direction so that we can write the dissolution flux approximately as
\begin{align}
 J_z = \frac{-D (\rho^{1}_{s}-\rho^{1}_{top})}{L_z},
 \label{eq:flux}
\end{align}
in which $\rho^{1}_{s}$ is the saturation density of liquid $1$ near the
droplet surface and $\rho^{1}_{top}$ is the density of liquid $1$ in the
ambient air. 

With~\eqnref{flux}, we obtain the rate of mass of a single droplet released
into the ambient air at the top surface as
\begin{align}
   dM/dt = L^2 J_z= \frac{-D L^2(\rho^{1}_{s}-\rho^{1}_{top})}{L_z},
   \label{eq:dmass2}
\end{align}
where $L^2$ represents the effective area of the top surface for the
corresponding single droplet. \eqnref{dmass2} indicates that the rate of mass
loss decreases with increasing layer thickness $L_z$, which is consistent with
the simulation results indicating that the droplet height has a slower
decreasing rate with increasing $L_z$ (see~\figref{htcr} and~\figref{htca}).
Moreover, based on~\eqnref{dmass2}, when the layer thickness is fixed, the rate
of mass loss can be treated as constant during the dissolution process. Thus,
the height of the droplet decreases faster when the droplet volume decreases,
which agrees qualitatively with the simulation results shown in~\figref{htcr}
and~\figref{htca}, i.e.  the decreasing rate of the droplet height increases
towards the end of the life of the dissolving droplet.
\revisedtext{We note that Eq.~14 is equivalent to the Noyes-Whitney equation~\cite{Smith2015, YANG2018}, 
which is generally used to describe the rate of a solute dissolving in a solvent.}

By integrating~\eqnref{dmass2}, we get the time evolution of the droplet mass as
\begin{align}
 M = M_0- \frac{D L^2 (\rho^{1}_{s}-\rho^{1}_{top})t}{L_z},
 \label{eq:masstime}
\end{align}
where $M_0$ is the initial mass of the droplet. 
From~\eqnref{masstime}, we can obtain the lifetime of the droplet as 
\begin{align}
 T =\frac{M_0L_z}{D L^2 (\rho^1_{s}-\rho^1_{top})}.
 \label{eq:time}
\end{align}
\eqnref{time} shows that the dissolution time is proportional to the length $L_z$, 
which is consistent with our simulation results shown in~\figref{scale}.

The total mass of a spherical cap-shaped droplet is
\begin{align}
 M = \frac{\pi \rho^{1}_{d}h}{6}(3R^2+h^2), 
 \label{eq:mass}
\end{align}
where $\rho^1_{d}$ is the density of liquid $1$ inside the droplet.
In the case of the droplet being dissolved in CA mode ($R=R_0$), we obtain
the time derivative of the total mass from~\eqnref{mass} as 
\begin{align}
 dM/dt = \frac{\pi \rho^{1}_{d}}{2}(R_{0}^{2}+h^2) dh/dt.
 \label{eq:dmass1}
\end{align}
By comparing~\eqnref{dmass1} and \eqnref{dmass2}, we reach
\begin{align}
 dh/dt = \frac{-2DL^2(\rho^{1}_{s}-\rho^{1}_{top})}{\pi\rho^{1}_{d}L_z}\frac{1}{R_{0}^{2}+h^2}.
 \label{eq:dhdt}
\end{align}
Eq.~\ref{eq:dhdt} is solved numerically using a 4th-order Runge-Kutta algorithm
and compared to the simulation results in  \figref{htcr} for different layer
thicknesses $L_z=72$, $144$, $288$, $576$.  Our theoretical model (solid lines)
captures the qualitative features of the time evolution of the droplet height
for all simulated systems and quantitatively agrees with the numerical results
for a thick liquid layer, i.e. $L_z\geq 288$ ($L_z/h_0\geq4$). \revisedtext{We note that the increasing layer thickness induces a higher
hydrostatic pressure, which may affect the stability of the surface droplets.
In experiments, stable surface droplets are formed~\cite{ZhangPNAS2015, Dietrich2015} when the ratio of layer
thickness and droplet height is of order $10^3$. This range is similar to our simulation parameters.  A detailed understanding of the effect of the hydrostatic pressure on the stability of the droplet calls for
a systematic experimental investigation and extensive theoretical analysis, which is beyond the scope of the current work. }

If the droplet dissolves in the CA mode ($\theta=\theta_a$), we can write its total mass as 
\begin{align}
 M = \frac{\pi \rho^{1}_{d} (3\sin\theta_a+\cos\theta_a-1) }{3(1-\cos\theta_a)} h^3
\end{align}
and then we obtain the time derivative of mass as 
\begin{align}
  dM/dt = \frac{\pi \rho^{1}_{d} (3\sin\theta_a+\cos\theta_a-1)  h^2}{(1-\cos\theta_a)}  dh/dt.
  \label{eq:dmassca}
\end{align}
By comparing~\eqnref{dmassca} with~\eqnref{dmass2}, we get 
\begin{align}
 dh/dt = \frac{-D L^2(\rho^{1}_{s}-\rho^{1}_{top})}{L_z\rho^{1}_{d} \pi } \frac{1-\cos\theta_a} {3\sin\theta_a+\cos\theta_a-1} \frac{1}{h^2}.
 \label{eq:dhdtca}
\end{align}
If the contact angle of the droplet is $\theta_a=90^{\circ}$, \eqnref{dhdtca} is simplified to
\begin{align}
 dh/dt = \frac{-D L^2(\rho^{1}_{s}-\rho^{1}_{top})}{2L_z\rho^{1}_{d} \pi}\frac{1}{h^2}.
 \label{eq:ca}
\end{align}
In~\figref{htca} we again compare the theoretical analysis~\eqnref{ca} with
simulation results for different layer thicknesses $L_z=72$, $144$, $288$,
$576$. As in the previous case, our theoretical model agrees qualitatively with the simulation results
for all the systems and we obtain quantitative agreement with simulation
results for very thick liquid layers, i.e. $L_z=576$ ($L_z/h_0\geq9$).

The good agreement between the theoretical model and our simulation results for both CR and CA modes 
indicates that the limiting diffusion process is at the top interface and 
the dynamics of the contact line of the droplet can be neglected for predicting 
the lifetime of dissolving surface droplet if $L_z \gg h_0$.
The theoretical model performs worse when the droplet dissolves in CA instead
of in CR mode.  A possible explanation is that our model is valid in the limit
of densely distributed droplets on the substrate $L/2R_0 \approx 1$ and it does
not consider the variation of the distance between neighboring droplets.  In CR
mode, the distance of neighboring droplets keeps constant, whereas it increases
in CA mode. This increase weakens the collective effect and induces a sharp
non-linear decrease of the density around the droplet surface (\figref{fca-4}).
The latter is less accurately described by the dissolution flux equation
(\eqnref{flux}) assuming a linear density gradient. 
We note that
our lattice Boltzmann simulations recover the unsteady diffusion equation.
Therefore, the good agreement between our simulation results and our
theoretical model also indicates that the quasistatic diffusion assumption is
valid for $L_z\gg h_0$.

\section{Conclusion}
We demonstrated that the thickness of the liquid layer surrounding
immersed and dissolving surface droplets strongly influences the dissolution and the lifetime of the droplets.
This holds if the density gradient of the droplet liquid is fully developed in the surrounding liquid.  
We performed $3D$ lattice Boltzmann simulations of the dissolution of droplets
in both constant radius and constant angle modes. 
In the near field of the droplet, we observed the convective effect introduced by neighboring droplets 
on the density gradient and the divergence of the dissolution flux near the contact line for small contact angles, 
which is consistent with experimental observations~\cite{Laghezza2016} and theoretical contributions~\cite{Popov2005}.
However, in the far field of the droplet, the density is homogeneous along the horizontal direction and 
its gradient stays almost constant during dissolution when the liquid layer thickness is much larger than the droplet height.
Additionally, in both modes, the
lifetime of the dissolving droplets increases approximately linearly with
increasing the thickness of the liquid layer.

We proposed a simple theoretical model assuming a quasistatic diffusion
equation in the limit of liquid layer thickness much larger than the droplet
height $L_z/h \gg 1$. Our model predicts that the rate of mass loss is a
linear function of layer thickness $L_z$, which confirms the simulation
results.
Moreover, our model qualitatively captures the time evolution of the droplet
height and agrees quantitatively with simulation results for thick liquid
layers. Surprisingly, even in the range where the layer thickness is similar
to the droplet height, $L_z \approx h$, the lifetime predicted by our
theoretical model is of the same order as the one obtained from the
simulations. Therefore, our model can be used to estimate the lifetime of
dissolving droplets quickly regardless of the thickness of the liquid layer and
the modes of contact line dynamics.

In a future work we plan to extend this study to non-regularly distributed 
surface droplets and droplets with polydisperse sizes~\cite{ZhangPNAS2015,BaoLei2016}.
It would be also interesting to investigate the effect of liquid layer thickness on the 
buoyancy-driven convective flow~\cite{Dietrich2016} when the two liquids 
have a large density difference. \revisedtext{Our lattice Boltzmann method can be directly applied to this problem because it recovers the Navier-Stokes and unsteady diffusion equations~\citep{DennisXieJens2016}.}

\begin{acknowledgments}
We thank D.\ Lohse and X.\ Zhang for fruitful discussions.
Financial support is acknowledged from the Netherlands Organization for
Scientific Research (NWO) through an NWO Industrial
Partnership Programme (IPP).  This research programme is co-financed by
Océ-Technologies B.V., University of Twente and Eindhoven University of
Technology. We thank the J\"ulich Supercomputing Centre and the High
Performance Computing Center Stuttgart for the technical support and allocated
CPU time.
 \end{acknowledgments}

 \providecommand*{\mcitethebibliography}{\thebibliography}
\csname @ifundefined\endcsname{endmcitethebibliography}
{\let\endmcitethebibliography\endthebibliography}{}


\begin{mcitethebibliography}{39}
\providecommand*{\natexlab}[1]{#1}
\providecommand*{\mciteSetBstSublistMode}[1]{}
\providecommand*{\mciteSetBstMaxWidthForm}[2]{}
\providecommand*{\mciteBstWouldAddEndPuncttrue}
  {\def\EndOfBibitem{\unskip.}}
\providecommand*{\mciteBstWouldAddEndPunctfalse}
  {\let\EndOfBibitem\relax}
\providecommand*{\mciteSetBstMidEndSepPunct}[3]{}
\providecommand*{\mciteSetBstSublistLabelBeginEnd}[3]{}
\providecommand*{\EndOfBibitem}{}
\mciteSetBstSublistMode{f}
\mciteSetBstMaxWidthForm{subitem}
{(\emph{\alph{mcitesubitemcount}})}
\mciteSetBstSublistLabelBeginEnd{\mcitemaxwidthsubitemform\space}
{\relax}{\relax}

\bibitem[Lohse and Zhang(2015)]{Detlef2015}
D.~Lohse and X.~Zhang, \emph{Rev. Mod. Phys.}, 2015, \textbf{87},
  981--1035\relax
\mciteBstWouldAddEndPuncttrue
\mciteSetBstMidEndSepPunct{\mcitedefaultmidpunct}
{\mcitedefaultendpunct}{\mcitedefaultseppunct}\relax
\EndOfBibitem
\bibitem[M{\'{e}}ndez-Vilas \emph{et~al.}(2009)M{\'{e}}ndez-Vilas,
  J{\'{o}}dar-Reyes, and Gonz{\'{a}}lez-Mart{\'{i}}n]{Mendez-Vilas2009}
A.~M{\'{e}}ndez-Vilas, A.~J{\'{o}}dar-Reyes and M.~L.
  Gonz{\'{a}}lez-Mart{\'{i}}n, \emph{Small}, 2009, \textbf{5}, 1366--1390\relax
\mciteBstWouldAddEndPuncttrue
\mciteSetBstMidEndSepPunct{\mcitedefaultmidpunct}
{\mcitedefaultendpunct}{\mcitedefaultseppunct}\relax
\EndOfBibitem
\bibitem[Chiu and Lorenz(2009)]{Chiu2009}
D.~T. Chiu and R.~M. Lorenz, \emph{Acc. Chem. Res.}, 2009, \textbf{42},
  649--658\relax
\mciteBstWouldAddEndPuncttrue
\mciteSetBstMidEndSepPunct{\mcitedefaultmidpunct}
{\mcitedefaultendpunct}{\mcitedefaultseppunct}\relax
\EndOfBibitem
\bibitem[Shemesh \emph{et~al.}(2014)Shemesh, {Ben Arye}, Avesar, Kang, Fine,
  Super, Meller, Ingber, and Levenberg]{Shemesh2014}
J.~Shemesh, T.~{Ben Arye}, J.~Avesar, J.~H. Kang, A.~Fine, M.~Super, A.~Meller,
  D.~E. Ingber and S.~Levenberg, \emph{Proc. Natl. Acad. Sci.}, 2014,
  \textbf{111}, 11293--11298\relax
\mciteBstWouldAddEndPuncttrue
\mciteSetBstMidEndSepPunct{\mcitedefaultmidpunct}
{\mcitedefaultendpunct}{\mcitedefaultseppunct}\relax
\EndOfBibitem
\bibitem[Zhang \emph{et~al.}(2015)Zhang, Lu, Tan, Bao, He, Sun, and
  Lohse]{ZhangPNAS2015}
X.~Zhang, Z.~Lu, H.~Tan, L.~Bao, Y.~He, C.~Sun and D.~Lohse, \emph{Proc. Natl.
  Acad. Sci.}, 2015, \textbf{112}, 9253--9257\relax
\mciteBstWouldAddEndPuncttrue
\mciteSetBstMidEndSepPunct{\mcitedefaultmidpunct}
{\mcitedefaultendpunct}{\mcitedefaultseppunct}\relax
\EndOfBibitem
\bibitem[Bao \emph{et~al.}(2016)Bao, Werbiuk, Lohse, and Zhang]{BaoLei2016}
L.~Bao, Z.~Werbiuk, D.~Lohse and X.~Zhang, \emph{J. Phys. Chem. Lett.}, 2016,
  \textbf{7}, 1055--1059\relax
\mciteBstWouldAddEndPuncttrue
\mciteSetBstMidEndSepPunct{\mcitedefaultmidpunct}
{\mcitedefaultendpunct}{\mcitedefaultseppunct}\relax
\EndOfBibitem
\bibitem[Dietrich \emph{et~al.}(2015)Dietrich, Kooij, Zhang, Zandvliet, and
  Lohse]{Dietrich2015}
E.~Dietrich, E.~S. Kooij, X.~Zhang, H.~J.~W. Zandvliet and D.~Lohse,
  \emph{Langmuir}, 2015, \textbf{31}, 4696--4703\relax
\mciteBstWouldAddEndPuncttrue
\mciteSetBstMidEndSepPunct{\mcitedefaultmidpunct}
{\mcitedefaultendpunct}{\mcitedefaultseppunct}\relax
\EndOfBibitem
\bibitem[Zhang and Ducker(2008)]{Xuehua2008}
X.~Zhang and W.~Ducker, \emph{Langmuir}, 2008, \textbf{24}, 110--115\relax
\mciteBstWouldAddEndPuncttrue
\mciteSetBstMidEndSepPunct{\mcitedefaultmidpunct}
{\mcitedefaultendpunct}{\mcitedefaultseppunct}\relax
\EndOfBibitem
\bibitem[Day \emph{et~al.}(2012)Day, Manz, and Zhang]{Day2012}
P.~Day, A.~Manz and Y.~Zhang, \emph{Microdroplet Technology Principles and
  Emerging Applications in Biology and Chemistry}, {Springer}, 2012\relax
\mciteBstWouldAddEndPuncttrue
\mciteSetBstMidEndSepPunct{\mcitedefaultmidpunct}
{\mcitedefaultendpunct}{\mcitedefaultseppunct}\relax
\EndOfBibitem
\bibitem[Zhu \emph{et~al.}(2018)Zhu, Verzicco, Zhang, and Lohse]{Xiaojue2018}
X.~Zhu, R.~Verzicco, X.~Zhang and D.~Lohse, \emph{Soft Matter}, 2018,
  \textbf{14}, 2006--2014\relax
\mciteBstWouldAddEndPuncttrue
\mciteSetBstMidEndSepPunct{\mcitedefaultmidpunct}
{\mcitedefaultendpunct}{\mcitedefaultseppunct}\relax
\EndOfBibitem
\bibitem[Michelin \emph{et~al.}(2018)Michelin, Gu\'erin, and Lauga]{Lauga2018}
S.~Michelin, E.~Gu\'erin and E.~Lauga, \emph{Phys. Rev. Fluids}, 2018,
  \textbf{3}, 043601\relax
\mciteBstWouldAddEndPuncttrue
\mciteSetBstMidEndSepPunct{\mcitedefaultmidpunct}
{\mcitedefaultendpunct}{\mcitedefaultseppunct}\relax
\EndOfBibitem
\bibitem[Cazabat and Gu{\'{e}}na(2010)]{Cazabat2010}
A.-M. Cazabat and G.~Gu{\'{e}}na, \emph{Soft Matter}, 2010, \textbf{6},
  2591--2612\relax
\mciteBstWouldAddEndPuncttrue
\mciteSetBstMidEndSepPunct{\mcitedefaultmidpunct}
{\mcitedefaultendpunct}{\mcitedefaultseppunct}\relax
\EndOfBibitem
\bibitem[Deegan \emph{et~al.}(1997)Deegan, Bakajin, Dupont, Huber, Nagel, and
  Witten]{Deegan1997}
R.~D. Deegan, O.~Bakajin, T.~F. Dupont, G.~Huber, S.~R. Nagel and T.~A. Witten,
  \emph{Nature}, 1997, \textbf{389}, 827--829\relax
\mciteBstWouldAddEndPuncttrue
\mciteSetBstMidEndSepPunct{\mcitedefaultmidpunct}
{\mcitedefaultendpunct}{\mcitedefaultseppunct}\relax
\EndOfBibitem
\bibitem[Deegan \emph{et~al.}(2000)Deegan, Bakajin, Dupont, Huber, Nagel, and
  Witten]{Deegan2000}
R.~D. Deegan, O.~Bakajin, T.~F. Dupont, G.~Huber, S.~R. Nagel and T.~A. Witten,
  \emph{Phys. Rev. E.}, 2000, \textbf{62}, 756--765\relax
\mciteBstWouldAddEndPuncttrue
\mciteSetBstMidEndSepPunct{\mcitedefaultmidpunct}
{\mcitedefaultendpunct}{\mcitedefaultseppunct}\relax
\EndOfBibitem
\bibitem[Popov(2005)]{Popov2005}
Y.~O. Popov, \emph{Phys. Rev. E}, 2005, \textbf{71}, 036313\relax
\mciteBstWouldAddEndPuncttrue
\mciteSetBstMidEndSepPunct{\mcitedefaultmidpunct}
{\mcitedefaultendpunct}{\mcitedefaultseppunct}\relax
\EndOfBibitem
\bibitem[Lebedev(1965)]{Lebedev}
N.~N. Lebedev, \emph{{Special Functions and Their Applications, re- vised
  English ed. }}, Prentice-Hall, 1965\relax
\mciteBstWouldAddEndPuncttrue
\mciteSetBstMidEndSepPunct{\mcitedefaultmidpunct}
{\mcitedefaultendpunct}{\mcitedefaultseppunct}\relax
\EndOfBibitem
\bibitem[Laghezza \emph{et~al.}(2016)Laghezza, Dietrich, Yeomans,
  Ledesma-Aguilar, Kooij, Zandvliet, and Lohse]{Laghezza2016}
G.~Laghezza, E.~Dietrich, J.~M. Yeomans, R.~Ledesma-Aguilar, E.~S. Kooij,
  H.~J.~W. Zandvliet and D.~Lohse, \emph{Soft Matter}, 2016, \textbf{12},
  5787--5796\relax
\mciteBstWouldAddEndPuncttrue
\mciteSetBstMidEndSepPunct{\mcitedefaultmidpunct}
{\mcitedefaultendpunct}{\mcitedefaultseppunct}\relax
\EndOfBibitem
\bibitem[Dietrich \emph{et~al.}(2016)Dietrich, Wildeman, Visser, Hofhuis,
  Kooij, Zandvliet, and Lohse]{Dietrich2016}
E.~Dietrich, S.~Wildeman, C.~W. Visser, K.~Hofhuis, E.~S. Kooij, H.~J.~W.
  Zandvliet and D.~Lohse, \emph{J. Fluid Mech.}, 2016, \textbf{794},
  45--67\relax
\mciteBstWouldAddEndPuncttrue
\mciteSetBstMidEndSepPunct{\mcitedefaultmidpunct}
{\mcitedefaultendpunct}{\mcitedefaultseppunct}\relax
\EndOfBibitem
\bibitem[Succi(2001)]{Succi2001}
S.~Succi, \emph{{The} {Lattice} {Boltzmann} {Equation}: {For} {Fluid}
  {Dynamics} and {Beyond}}, Oxford University Press, 2001\relax
\mciteBstWouldAddEndPuncttrue
\mciteSetBstMidEndSepPunct{\mcitedefaultmidpunct}
{\mcitedefaultendpunct}{\mcitedefaultseppunct}\relax
\EndOfBibitem
\bibitem[Shan and Chen(1993)]{Shan1993}
X.~Shan and H.~Chen, \emph{Phys. Rev. E}, 1993, \textbf{47}, 1815\relax
\mciteBstWouldAddEndPuncttrue
\mciteSetBstMidEndSepPunct{\mcitedefaultmidpunct}
{\mcitedefaultendpunct}{\mcitedefaultseppunct}\relax
\EndOfBibitem
\bibitem[Liu \emph{et~al.}(2016)Liu, Kang, Leonardi, Schmieschek,
  Narv{\'{a}}ez, Jones, Williams, Valocchi, and Harting]{Liu2016}
H.~Liu, Q.~Kang, C.~R. Leonardi, S.~Schmieschek, A.~Narv{\'{a}}ez, B.~D. Jones,
  J.~R. Williams, A.~J. Valocchi and J.~Harting, \emph{Computat. Geosci.},
  2016, \textbf{20}, 777--805\relax
\mciteBstWouldAddEndPuncttrue
\mciteSetBstMidEndSepPunct{\mcitedefaultmidpunct}
{\mcitedefaultendpunct}{\mcitedefaultseppunct}\relax
\EndOfBibitem
\bibitem[Chen \emph{et~al.}(2014)Chen, Kang, Mu, He, and Tao]{CHEN2014210}
L.~Chen, Q.~Kang, Y.~Mu, Y.-L. He and W.-Q. Tao, \emph{International Journal of
  Heat and Mass Transfer}, 2014, \textbf{76}, 210 -- 236\relax
\mciteBstWouldAddEndPuncttrue
\mciteSetBstMidEndSepPunct{\mcitedefaultmidpunct}
{\mcitedefaultendpunct}{\mcitedefaultseppunct}\relax
\EndOfBibitem
\bibitem[Ledesma-Aguilar \emph{et~al.}(2014)Ledesma-Aguilar, Vella, and
  Yeomans]{LedVelYeo14}
R.~Ledesma-Aguilar, D.~Vella and J.~M. Yeomans, \emph{Soft Matter}, 2014,
  \textbf{10}, 8267\relax
\mciteBstWouldAddEndPuncttrue
\mciteSetBstMidEndSepPunct{\mcitedefaultmidpunct}
{\mcitedefaultendpunct}{\mcitedefaultseppunct}\relax
\EndOfBibitem
\bibitem[Laghezza \emph{et~al.}(2016)Laghezza, Dietrich, Yeomans,
  Ledesma-Aguilar, Kooij, Zandvliet, and Lohse]{AguVelYeo16}
G.~Laghezza, E.~Dietrich, J.~M. Yeomans, R.~Ledesma-Aguilar, E.~S. Kooij,
  H.~J.~W. Zandvliet and D.~Lohse, \emph{Soft Matter}, 2016, \textbf{12},
  5787\relax
\mciteBstWouldAddEndPuncttrue
\mciteSetBstMidEndSepPunct{\mcitedefaultmidpunct}
{\mcitedefaultendpunct}{\mcitedefaultseppunct}\relax
\EndOfBibitem
\bibitem[Jansen \emph{et~al.}(2013)Jansen, Sotthewes, van Swigchem, Zandvliet,
  and Kooij]{Jansen2013}
H.~P. Jansen, K.~Sotthewes, J.~van Swigchem, H.~J.~W. Zandvliet and E.~S.
  Kooij, \emph{Phys. Rev. E}, 2013, \textbf{88}, 013008\relax
\mciteBstWouldAddEndPuncttrue
\mciteSetBstMidEndSepPunct{\mcitedefaultmidpunct}
{\mcitedefaultendpunct}{\mcitedefaultseppunct}\relax
\EndOfBibitem
\bibitem[Hessling \emph{et~al.}(2017)Hessling, Xie, and
  Harting]{DennisXieJens2016}
D.~Hessling, Q.~Xie and J.~Harting, \emph{J. Chem. Phys.}, 2017, \textbf{146},
  054111\relax
\mciteBstWouldAddEndPuncttrue
\mciteSetBstMidEndSepPunct{\mcitedefaultmidpunct}
{\mcitedefaultendpunct}{\mcitedefaultseppunct}\relax
\EndOfBibitem
\bibitem[Xie and Harting(2018)]{XH18a}
Q.~Xie and J.~Harting, \emph{Langmuir}, 2018, \textbf{34}, 5303--5311\relax
\mciteBstWouldAddEndPuncttrue
\mciteSetBstMidEndSepPunct{\mcitedefaultmidpunct}
{\mcitedefaultendpunct}{\mcitedefaultseppunct}\relax
\EndOfBibitem
\bibitem[Hyv\"aluoma \emph{et~al.}(2011)Hyv\"aluoma, Kunert, and
  Harting]{HyvKunHarH11}
J.~Hyv\"aluoma, C.~Kunert and J.~Harting, \emph{J. Phys. Condens. Matter},
  2011, \textbf{23}, 184106\relax
\mciteBstWouldAddEndPuncttrue
\mciteSetBstMidEndSepPunct{\mcitedefaultmidpunct}
{\mcitedefaultendpunct}{\mcitedefaultseppunct}\relax
\EndOfBibitem
\bibitem[Jansen and Harting(2011)]{Jansen2011}
F.~Jansen and J.~Harting, \emph{Phys. Rev. E}, 2011, \textbf{83}, 046707\relax
\mciteBstWouldAddEndPuncttrue
\mciteSetBstMidEndSepPunct{\mcitedefaultmidpunct}
{\mcitedefaultendpunct}{\mcitedefaultseppunct}\relax
\EndOfBibitem
\bibitem[Frijters \emph{et~al.}(2012)Frijters, G\"{u}nther, and
  Harting]{Frijters2012}
S.~Frijters, F.~G\"{u}nther and J.~Harting, \emph{Soft Matter}, 2012,
  \textbf{8}, 6542--6556\relax
\mciteBstWouldAddEndPuncttrue
\mciteSetBstMidEndSepPunct{\mcitedefaultmidpunct}
{\mcitedefaultendpunct}{\mcitedefaultseppunct}\relax
\EndOfBibitem
\bibitem[G\"{u}nther \emph{et~al.}(2014)G\"{u}nther, Frijters, and
  Harting]{Gunther2013}
F.~G\"{u}nther, S.~Frijters and J.~Harting, \emph{Soft Matter}, 2014,
  \textbf{10}, 4977\relax
\mciteBstWouldAddEndPuncttrue
\mciteSetBstMidEndSepPunct{\mcitedefaultmidpunct}
{\mcitedefaultendpunct}{\mcitedefaultseppunct}\relax
\EndOfBibitem
\bibitem[Xie \emph{et~al.}(2016)Xie, Davies, and Harting]{Xie2016}
Q.~Xie, G.~B. Davies and J.~Harting, \emph{Soft Matter}, 2016, \textbf{12},
  6566--6574\relax
\mciteBstWouldAddEndPuncttrue
\mciteSetBstMidEndSepPunct{\mcitedefaultmidpunct}
{\mcitedefaultendpunct}{\mcitedefaultseppunct}\relax
\EndOfBibitem
\bibitem[Qian \emph{et~al.}(1992)Qian, D'Humi\`{e}res, and Lallemand]{Qian1992}
Y.~H. Qian, D.~D'Humi\`{e}res and P.~Lallemand, \emph{Europhys. Lett.}, 1992,
  \textbf{17}, 479--484\relax
\mciteBstWouldAddEndPuncttrue
\mciteSetBstMidEndSepPunct{\mcitedefaultmidpunct}
{\mcitedefaultendpunct}{\mcitedefaultseppunct}\relax
\EndOfBibitem
\bibitem[Krüger \emph{et~al.}(2016)Krüger, Kusumaatmaja, Kuzmin, Shardt,
  Silva, and Viggen]{Timm2016}
T.~Krüger, H.~Kusumaatmaja, A.~Kuzmin, O.~Shardt, G.~Silva and E.~Viggen,
  \emph{{The Lattice Boltzmann Method - Principles and Practice}}, Springer,
  2016\relax
\mciteBstWouldAddEndPuncttrue
\mciteSetBstMidEndSepPunct{\mcitedefaultmidpunct}
{\mcitedefaultendpunct}{\mcitedefaultseppunct}\relax
\EndOfBibitem
\bibitem[Huang \emph{et~al.}(2007)Huang, Thorne, Schaap, and Sukop]{Huang2007}
H.~Huang, D.~T. Thorne, M.~G. Schaap and M.~C. Sukop, \emph{Phys. Rev. E},
  2007, \textbf{76}, 066701\relax
\mciteBstWouldAddEndPuncttrue
\mciteSetBstMidEndSepPunct{\mcitedefaultmidpunct}
{\mcitedefaultendpunct}{\mcitedefaultseppunct}\relax
\EndOfBibitem
\bibitem[Bao \emph{et~al.}(2018)Bao, Spandan, Yang, Dyett, Verzicco, Lohse, and
  Zhang]{Lei2018}
L.~Bao, V.~Spandan, Y.~Yang, B.~Dyett, R.~Verzicco, D.~Lohse and X.~Zhang,
  \emph{Lab Chip}, 2018, \textbf{18}, 1066--1074\relax
\mciteBstWouldAddEndPuncttrue
\mciteSetBstMidEndSepPunct{\mcitedefaultmidpunct}
{\mcitedefaultendpunct}{\mcitedefaultseppunct}\relax
\EndOfBibitem
\bibitem[Yang \emph{et~al.}(2019)Yang, Yuan, and Zhao]{Zhao2019}
J.~Yang, Q.~Yuan and Y.~Zhao, \emph{Science China Physics, Mechanics} \&\emph{Astronomy}, 2019, \textbf{62}, 124611\relax
\mciteBstWouldAddEndPuncttrue
\mciteSetBstMidEndSepPunct{\mcitedefaultmidpunct}
{\mcitedefaultendpunct}{\mcitedefaultseppunct}\relax
\EndOfBibitem
\bibitem[Yang \emph{et~al.}(2018)Yang, Yuan, and Zhao]{YANG2018}
J.~Yang, Q.~Yuan and Y.~Zhao, \emph{Int. J. Heat Mass Transf.}, 2018,
  \textbf{118}, 201--207\relax
\mciteBstWouldAddEndPuncttrue
\mciteSetBstMidEndSepPunct{\mcitedefaultmidpunct}
{\mcitedefaultendpunct}{\mcitedefaultseppunct}\relax
\EndOfBibitem
\bibitem[Smith(2015)]{Smith2015}
B.~T. Smith, \emph{{Remington Education: Physical Pharmacy}}, Pharmaceutical
  Press, 2015\relax
\mciteBstWouldAddEndPuncttrue
\mciteSetBstMidEndSepPunct{\mcitedefaultmidpunct}
{\mcitedefaultendpunct}{\mcitedefaultseppunct}\relax
\EndOfBibitem
\end{mcitethebibliography}
\end{document}